\documentclass{article}
\usepackage[T1]{fontenc} % add special characters (e.g., umlaute)
\usepackage[utf8]{inputenc} % set utf-8 as default input encoding
\usepackage{templates/ISMIR/ismir,amsmath,cite,url}
\usepackage{graphicx}
\usepackage{style,myismir}
\usepackage{multirow}
\usepackage{booktabs}
\usepackage{lipsum}
\usepackage[dvipsnames]{xcolor}
\usepackage{xfrac}
\usepackage{caption}
\usepackage{subcaption}
\usepackage{enumitem}
\usepackage{xfrac}
\usepackage{tikz}
\usepackage{balance}
\usepackage{xcolor}
\usepackage{diagbox}

\usepackage{xfp}

\usetikzlibrary{calc}

\title{A Model You Can Hear:\\Audio Identification with Playable Prototypes}

\author{Romain Loiseau\textsuperscript{1, 2}
\and
Baptiste Bouvier\textsuperscript{3}
\and
Yann Teytaut\textsuperscript{3}
\and
Elliot Vincent\textsuperscript{1,4}
\and
Mathieu Aubry\textsuperscript{1}
\and
Loic Landrieu\textsuperscript{2}
\and
{{\textsuperscript{1}LIGM, Ecole des Ponts, Univ Gustave Eiffel, CNRS, France}}\\
{{\textsuperscript{2}LASTIG, Univ Gustave Eiffel, IGN, ENSG}}\\
{{\textsuperscript{3}STMS Lab, UMR 9912 (IRCAM, CNRS, Sorbonne University), Paris, France}}\\
{{\textsuperscript{4}INRIA and DIENS (ENS-PSL, CNRS, INRIA)}}
}

\multauthor
{Romain Loiseau$^{1,2}$ \hspace{1cm} Baptiste Bouvier$^3$ \hspace{1cm} Yann Teytaut$^3$}
{\bfseries{~~~Elliot Vincent$^{1,4}$~ \hspace{1cm} ~~Mathieu Aubry$^1$~ \hspace{1cm} Loic Landrieu$^2$}\\
$^1$ LIGM, Ecole des Ponts, Univ Gustave Eiffel, CNRS, France\\
$^2$ LASTIG, Univ Gustave Eiffel, IGN, ENSG\\
$^3$ STMS Lab, UMR 9912 (IRCAM, CNRS, Sorbonne University), Paris, France\\
$^4$ INRIA and DIENS (ENS-PSL, CNRS, INRIA)\\
{\tt\small romain.loiseau@enpc.fr}
}

\def\authorname{R. Loiseau, B. Bouvier, Y. Teytaut, E. Vincent, M. Aubry, and L. Landrieu}

\begin{document}
\definecolor{Kselect}{RGB}{0,202,42}
\definecolor{Knotselect}{RGB}{203, 0, 0}
\definecolor{input}{RGB}{0,102,199}

\maketitle
\begin{abstract}
   Machine learning techniques have proved useful for classifying and analyzing audio content. However, recent methods typically rely on abstract and high-dimensional representations that are difficult to interpret. Inspired by transformation-invariant approaches developed for image and 3D data, we propose an audio identification model based on learnable spectral prototypes. Equipped with dedicated transformation networks, these prototypes can be used to cluster and classify input audio samples from large collections of sounds. Our model can be trained with or without supervision and reaches state-of-the-art results for speaker and instrument identification, while remaining easily interpretable. The code is available at: {\tt{\url{https://github.com/romainloiseau/a-model-you-can-hear}}}
\end{abstract}

\section{Introduction}

The emergence of deep learning approaches dedicated to audio analysis has led to significant performance improvements~\cite{abesser2020review, ndou2021music}.
Although these methods take sound excerpts as input, they typically rely on high-dimensional latent spaces, making the interpretation of their decisions difficult and limiting the insights gained on the considered data. Furthermore, such methods typically require large corpora of annotated sounds for supervision.
We take a different approach, inspired by the recent Deep Transformation-Invariant~(DTI) clustering~\cite{monnier2020deep} framework which learns prototypes in input space to analyze images or 3D point clouds~\cite{loiseau2021representing}.
Each prototype is associated with a set of transformations (\eg rotations or morphological transformations) learned in the manner of Spatial Transformation Networks~\cite{jaderberg2015spatial}. The DTI clustering model optimizes a reconstruction loss and associates each input with the prototype that provides the best reconstruction. We propose adapting this approach to the audio domain. The prototypes become log-scaled, Mel-frequency spectrograms, and their associated learnable transformations correspond to plausible sound alterations.
We also introduce an additional loss based on cross-entropy in the supervised setting, which we demonstrate can boost the classification accuracy while preserving some interpretability.

\begin{figure}[t]
    \centering
    \begin{center}
    \begin{subfigure}[t]{\linewidth}
        \begin{tabular}{@{}p{.35\linewidth}p{.61\linewidth}@{}}
            \teaserspect{2_Bassoons_Bassoon}{Bassoon}{sol}
            \teaserspect{12_Horns_Horn}{Horn}{sol}
            \teaserspect{18_Strings_Violin}{Violin}{sol}
            \teaserspect{20_Strings_Contrabass}{Contrabass}{sol}
            \teaserspect{29_Keyboards_Accordion}{Accordion}{sol}
        \end{tabular}
        \subcaption{Instrument dataset SOL~\cite{ballet1999studio,cella2020orchideasol}}
    \end{subfigure}
    \begin{subfigure}[t]{\linewidth}
        \begin{tabular}{@{}p{.35\linewidth}p{.61\linewidth}@{}}
            \teaserspect{78_3546_F_Jeannie}{Jeannie (F)}{libri}
            \teaserspect{88_224_F_CaitlinKelly}{Caitlin Kelly (F)}{libri}
            \teaserspect{102_6538_M_JuanFederico}{Juan Federico (M)}{libri}
            \teaserspect{117_1336_M_CharlieBlakemore}{Charlie (M)}{libri}
            \teaserspect{122_7688_M_LeslieWalden}{Leslie Walden (M)}{libri}
        \end{tabular}
        \subcaption{Speech dataset Librispeech~\cite{panayotov2015librispeech}}
    \end{subfigure}
    \end{center}
    \vspace{-1em}
    \caption{
    \textbf{Playable Prototypes.}~Our model rely on a small set of spectral prototypes that can be used for clustering and classification. We show examples of such prototypes learned from two datasets.
    }
    \label{fig:teaser}
\end{figure}
\section{Related Work}

\begin{figure*}
    \centering
    \includegraphics[trim={0 0 1.8cm 0},clip,width=.9\textwidth]{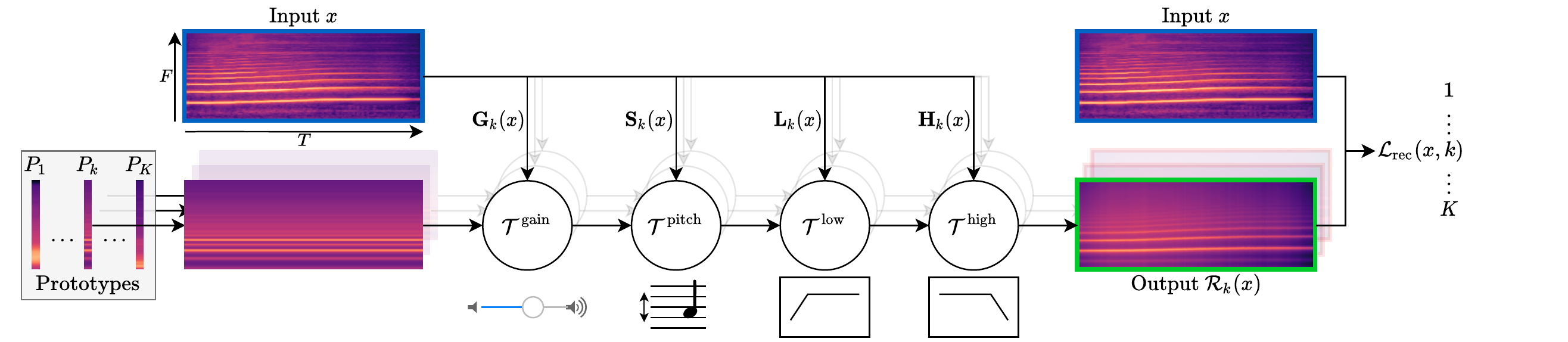}
    \caption{\textbf{Method overview.}~Given an \textcolor{input}{\textbf{input sound}}, we predict for each prototype a \emph{gain}, a \emph{pitch} shift, as  well as \textit{low} and \textit{high frequency filters} at each timestamp to generate the \textcolor{Kselect}{\textbf{output}}. Prototypes and transformations are learned jointly using a reconstruction loss in either a supervised or unsupervised setting.
    }
    \label{fig:pipeline}
\end{figure*}

\paragraph{Audio Classification.}Musical instrument and speaker identification are two of the standard tasks used to benchmark audio classification models.
Although early musical instrument identification methods focused on distinguishing individual tones achieve appreciable precision, support vector machines operating on spectro-temporal sound representations can reach almost perfect accuracy. Similarly to supervised classification of instruments that play individual notes that could be considered solved~\cite{bhalke2016automatic,lostanlen2018extended}, speaker identification is mostly handled by convolutional and/or recurrent neural networks~\cite{bai2021speaker}, which achieve almost perfect performances~\cite{ye2021deep}.
However, these models rely on complex and abstract latent representations that are not easily %, if at all,
interpretable and cannot be visualized as spectrograms or heard in the audio domain.

\paragraph{Prototype-Based Methods.} Auto-Encoders~\cite{tits2019visualization, roche2018autoencoders} learn a compact latent representation of an input sample and are trained using a reconstruction loss. Using regularizations and constraints, the latent space can be adjusted for various tasks, such as classification~\cite{naranjo2020open}. However, the learned features remain largely non-interpretable and abstract. Inspired by the literature on images~\cite{li2018deep}, the authors of~\cite{zinemanas2021deep} propose to generate prototypes in the latent space and analyze their similarity to the encoded input samples using a frequency-dependent measure. Although the latent prototypes can be decoded into input space, their role in class prediction remains obfuscated, as all similarities are mixed with a learned linear layer.

\paragraph{Transformation-Invariant Modeling.}
Deep transfor-mation-invariant clustering~\cite{monnier2020deep} learns explicit prototypes in input space. Each prototype is equipped with dedicated transformation networks, allowing a small set of prototypes to faithfully represent a collection of samples.
The resulting models can be used for downstream tasks such as classification~\cite{monnier2020deep}, few-shot segmentation~\cite{loiseau2021representing}, and even multi-object instance discovery~\cite{monnier2021dtisprites}. Jaderberg~\etal~\cite{jaderberg2015spatial} also proposes learning differentiable transformations in the input space, and feed the transformed data to a classification network. We propose to extend these ideas to the audio domain by learning prototypical log-Mel spectrograms along with adapted transformations.
\section{Method}

We consider a set of $N$ audio clips $x_1,\cdots,x_N\in\mathbb{R}^{F\times T}$, each characterized by their log-Mel spectrogram with $T$ time steps and a spectral resolution of $F$ bins. The samples are annotated with class labels $y_1, \cdots, y_N\in\mathcal{Y}$ with $\mathcal{Y}$ a given class set $\{1,\cdots,K\}$.
Our goal is to learn an interpretable model that can perform classification and clustering for the audio sample collection $x_1,\cdots,x_N$. Our model is based on prototypical spectrograms equipped with spectral transformation networks, which we present in \Secref{sec:dti}. We propose an unsupervised training scheme in \Secref{sec:unsupervised}, and a supervised setting in \Secref{sec:supervised}. Finally, we provide details on parameterization and training of the model in \Secref{sec:param}.

\subsection{{Sound Reconstruction Model}}\label{sec:dti}
Our model consists of $K$ prototypes $P_k\in\mathbb{R}^F$ directly interpretable in the spectral domain, each equipped with a set of dedicated transformation networks (See \figref{fig:pipeline}). Both prototypes and networks are jointly trained to reconstruct input samples. We can then use the reconstruction error as a measure of the compatibility between a sample and a given prototype.

\paragraph{Spectral Transformations Model.}
Even a single instrument or speaker cannot be meaningfully characterized by a single sound. In fact, they are typically capable of producing a rich and varied set of sounds. We propose equipping each prototype with a set of transformation networks that learn to change specific characteristics of the prototype, such as its amplitude and pitch, so that it can faithfully approximate a wide variety of samples. This also allows the prototypes to learn more subtle audio attributes, such as timbre or intonations.

Transformation networks associated to each prototype $P_k$ take as input an audio sample $x\in\mathbb{R}^{F \times T}$ as input and predicts a spectral transformation for each time step $t$ in $[1,T]$. For each prototype $k$, we propose a set of transformations specifically designed for the audio domain, illustrated in \figref{fig:visutransfo}:
\begin{figure*}
    \centering
    \input{figures/deformation}
    \caption{\textbf{Spectral Transformations.}~
    Our proposed transformations learn to alter the gain~\Subref{fig:visutransfo:gain}, pitch~\Subref{fig:visutransfo:pitch} and frequency support~(\subref{fig:visutransfo:lf},\subref{fig:visutransfo:hf}) of a given prototype spectrogram $P_k$.
    %The changes are exaggerated for the sake of illustration.
    }
    \label{fig:visutransfo}
\end{figure*}

\begin{itemize}[itemsep=0em, wide, labelwidth=!, labelindent=0pt, topsep=.5em]
\item\textit{Amplification.}~Given an audio sample $x$, we predict a gain $\mathbf{G}_k(x)\in\mathbb{R}^{T}$ for all time step $t$. The transformation $\mathcal{T}^\text{gain}_{g}(m)$ adds the gain $g$ to all frequencies of a log-Mel spectrogram $m \in \mathbb{R}^F$. 
\item\textit{Pitch Shifting.}~ 
We map an audio sample $x$ to a pitch shift $\mathbf{S}_k(x)\in\mathbb{R}^{T}$ for all time step $t$. The transformation $\mathcal{T}^\text{pitch}_{s}(m)$ shifts the log-Mel spectrogram $m\in\mathbb{R}^F$ %the
by $s$. We use linear interpolation for handling non-integer shifts.
\item\textit{High- and Low-Frequency Filter.}~
Given an audio sample $x$, we define for each time $t$ an affine low-frequency filter $\mathbf{L}_k(x)[t] \in \mathbf{R}^F$ by predicting its slope and cutoff frequency. Similarly, we predict a high-frequency filter $\mathbf{H}_k(x)[t] \in \mathbb{R}^F$.
The transformations $\mathcal{T}^\text{low}_L$ and $\mathcal{T}^\text{high}_H$ simply add the corresponding filters $L$ and $H$ to a log-Mel spectrogram.
\end{itemize}

\noindent These transformations can be written as follows for a log-Mel spectrogram $m \in \mathbb{R}^F$ and a mel-frequency $f \in [1,F]$:
\vspace{-1em}
\begin{align}
    \mathcal{T}^\text{gain}_g\left(m\right)[f]  &= m[f] + g\\
    \mathcal{T}^\text{pitch}_s\left(m\right)[f] &= m\left[\Phi(f, s)\right]\\
    \mathcal{T}^\text{low}_L\left(m\right)[f]   &= m[f] + L[f]\\
    \mathcal{T}^\text{high}_H\left(m\right)[f]  &= m[f] + H[f]~,
\end{align}
with 
%\vspace{-1em}
%\begin{align}
$
\Phi(f, s) = A \log_{10}\left(1 + s\left(10^{\sfrac{f}{A}} - 1\right)\right)
$
the mel-aware pitch shifting, with $A=2595$~\cite{stevens1937scale}. Our transformations have links with moment matching methods~\cite{tamaru2019generative, andreux2018music}. In fact, modulating the amplitude, pitch, and spectral support of spectrograms is similar to matching the moment of zeroth order (amplitude gain), first order (pitch shift), and second order (frequency filters).

\paragraph{Reconstruction Model.}
{For each prototype $k$ and each time step $t$, we successively apply the gain, pitch shift, low-pass, and high-pass transformations in this specific order. The resulting reconstruction $\mathcal{R}_k(x)$ of an input sound $x$ by a prototype $P_k$ is defined as follows:}

\vspace{-1em}
\begin{align}\label{eq:reconstruction}\nonumber
     \mathcal{R}_k(x)[t]=\;
     &
     \mathcal{T}^\text{high}_{\mathbf{H}_k(x)[t]}
     \circ
     \mathcal{T}^\text{low}_{\mathbf{L}_k(x)[t]}\\
     \circ\;&
     \mathcal{T}^\text{pitch}_{\mathbf{S}_k(x)[t]}
      \circ
    \mathcal{T}^\text{gain}_{\mathbf{G}_k(x)[t]}(P_k)~,
\end{align}
{where $\mathcal{R}_k(x)[t]$ denotes the $t$-th timestamp of the reconstruction, and $\circ$ denotes the composition of transformations.}
Note that the chosen transformations are constrained and limited on purpose, as we want each prototype to represent a restricted and consistent set of audio samples.

We measure the quality of the reconstruction $\mathcal{R}_k(x)$ of an input sample $x$ by the $k$-th prototype as the spectro-temporal average of their $\ell_2$ distance: 

\vspace{-1em}
\begin{align}
    \lrec\left(x,k\right) =
    \frac1T
    \sum_{t = 1}^T
    \left\Vert
        x[t] - \mathcal{R}_{k}(x)[t]
    \right\Vert^2~.
\end{align}
Note that since this function is differentiable and continuous, it can be used as a loss function to train prototypes and transformation networks as defined in the next sections.

\subsection{Unsupervised Training}\label{sec:unsupervised}

We propose an unsupervised learning scheme in which we do not have access to labels $y_1, \cdots, y_N$ during training. Our goal is to cluster the samples $x_1, \cdots, x_N$ into meaningful groups of sounds that can be well approximated by the same prototype spectrogram and its associated transformations. We can train our model by minimizing the following loss:

\vspace{-1em}\begin{align}
    \lclu=\sum_{n=1}^N\min_{k=1}^K\lrec(x_n,k)~.
\end{align}
This loss is the sum of the reconstruction errors with optimal cluster assignment and is a classical clustering objective used by K-means-based methods.

\begin{table*}[t]
    \centering
    
    \begin{tabular*}{\linewidth}{@{}l@{\extracolsep{\fill}}*{12}{c}@{}}
        \toprule
        & \multicolumn{6}{c}{SOL~\cite{ballet1999studio, cella2020orchideasol}}& \multicolumn{6}{c}{LibriSpeech~\cite{panayotov2015librispeech}} \\\cmidrule(lr){2-7}\cmidrule(lr){8-13}
        & \multicolumn{3}{c}{w/o supervision} & \multicolumn{3}{c}{w supervision}& \multicolumn{3}{c}{w/o supervision} & \multicolumn{3}{c}{w supervision} \\\cmidrule(lr){2-4}\cmidrule(lr){8-10}\cmidrule(lr){5-7}\cmidrule(lr){11-13}
        & OA & AA & $\lrec$ & OA & AA & $\lrec$ & OA & AA & $\lrec$ & OA & AA & $\lrec$ \\
        \midrule
        Raw prototypes                               & $29.3$	& $13.2$ & $12.4$ & $28.4$ & $39.5$	& $12.9$ & $13.3$ & $13.2$ & $8.6$ & $60.5$ & $59.5$ & $8.6$ \\
        ~~$\llcorner$~\emph{gain} transformation     & $32.2$	& $15.8$ & $3.4$  & $37.4$ & $40.1$	& $3.8$  & $44.4$ & $43.3$ & $3.4$ & $78.7$ & $79.1$ & $3.4$ \\
        ~~~~$\llcorner$~\emph{pitch-shifting}        & $\mathbf{37.6}$	& $\mathbf{19.1}$ & $2.6$  & $51.6$ & $46.1$	& $2.9$  & $46.8$ & $46.9$ & $2.6$ & $77.2$ & $77.8$ & $2.6$ \\
        ~~~~~~$\llcorner$~\emph{frequency-filters}   & $33.4$	& $14.6$ & $\mathbf{1.5}$  & $55.1$ & $47.7$	& $\mathbf{2.1}$  & $\mathbf{48.8}$ & $\mathbf{49.5}$ & $\mathbf{1.6}$ & $88.8$ & $89.4$ & $\mathbf{1.6}$ \\
	    ~~~~~~~~$\llcorner$~cross-entropy loss       & ---  & ---  & ---        & $\mathbf{98.9}$	& $\mathbf{94.5}$ & $2.5$  & ---	 & ---  & ---              & $\mathbf{99.9}$ & $\mathbf{99.9}$ & $1.9$ \\
        \bottomrule
    \end{tabular*}
    \caption{\textbf{Ablation Study.}~We show the impact of the increasing complexity of our modeling on datasets' validation sets.
    }
    \label{tab:ablation}
    \vspace{-1em}
\end{table*}

\subsection{Supervised Training}\label{sec:supervised}

In the supervised setting, each prototype and associated transformation networks are dedicated to a class and trained to reconstruct the samples from this class only. Once trained, the model predicts for an input $x$ the class of the prototype with the best reconstruction.

To promote better class discrimination, we propose a dual reconstruction-prediction objective. We associate each input $x$ with a probabilistic class prediction defined as the softmax of the negative reconstruction error between prototypes. 
For an input $x$ of label $y$, we define the prediction loss $\lce$:

\vspace{-1em}
\begin{align}
    \lce\left(x, y\right) =
    - \log \left(
    \frac{
    \exp\left({-\beta\lrec\left(x, y\right)}\right)
    }{
    \sum_{k=1}^K \exp\left({-\beta\lrec\left(x, k\right)}\right)
    }
    \right)
    ~,
\end{align}
where $\beta$ is a learnable parameter. %corresponding to the inverse temperature in the softmax.
This loss encourages each prototype to specialize in reconstructing the samples of its associated class. To train our network, we use a weighted sum of both losses:

\vspace{-1em}\begin{align}
    \lclass=\sum_{n=1}^N \lrec(x_n,y_n) + \lambda_\text{ce}\lce (x_n,y_n)~,
\end{align}
with $\lambda_\text{ce}$ an hyperparameter set to $0.01$.

\begin{table}[t]
    \centering
    \begin{center}
    \begin{tabular}{@{}lcc@{}}
        \toprule
        & OA & AA  \\
        \midrule
        \textbf{SOL}~\cite{ballet1999studio, cella2020orchideasol} \\
        ~~Autoencoder + K-means~\cite{bottou1995convergence}                    & $28.7$ & $12.3$\\
        ~~$\dagger$ APNet~\cite{zinemanas2021deep} + K-means~\cite{bottou1995convergence} & $\mathbf{37.3}$ & $\mathbf{18.2}$\\
        ~~Ours w/o supervision                                                  & $34.5$ & $15.4$\\
        \midrule
        \textbf{LibriSpeech}~\cite{panayotov2015librispeech} \\
        ~~Autoencoder + K-means~\cite{bottou1995convergence}                    & $11.0$ & $11.1$\\
        ~~$\dagger$  APNet~\cite{zinemanas2021deep} + K-means~\cite{bottou1995convergence} & $36.3$ & $36.4$\\
        ~~Ours w/o supervision                                                  & $\mathbf{48.6}$ & $\mathbf{49.5}$\\
        \bottomrule
    \end{tabular}
    \end{center}
    \vspace{-1em}
    \caption{\textbf{Clustering Results.}~Clustering performances on the test sets. $\dagger$:~{\small{Note that APNet requires labels at training time.}}}
    \label{tab:clustering}
    \vspace{-1em}
\end{table}

\subsection{Parameterization and Training Details}\label{sec:param}

\paragraph{Architecture.}
We implement the functions $\mathbf{G}_k$, $\mathbf{S}_k$, $\mathbf{L}_k$ and $\mathbf{H}_k$ as U-Net style neural networks~\cite{ronneberger2015u} operating on the log-Mel spectrograms $x\in\mathbb{R}^{T\times F}$.
To save parameters, all networks share the same encoder, defined as a sequence of 2-dimensional spectro-temporal convolutions and pooling operations. This encoder produces a sequence of feature maps $z_0,\cdots,z_R$ with decreasing temporal and spectral resolutions $T/2^r\times F/2^r$. The spectral dimension of these maps is then collapsed using learned pooling operations: $\tilde{z}_r$ is of resolution $T/2^r$.
Each decoder is then defined as a sequence of 1-dimensional temporal convolutions and unpooling operations. The spectral maps $\tilde{z}_0,\cdots,\tilde{z}_R$ are used through skip-connections in the decoders. The prototypes $P_1,\cdots,P_K$ and the inverse temperature $\beta$ are directly trainable parameters of the model.

\paragraph{Curriculum Learning.}
We ensure the stability of our model by gradually increasing its complexity along the training procedure. First, we learn the raw prototypes without any transformation. Once the training loss does not decrease for 10 straight epochs, we add the gain transformation. We then add the pitch shift and spectral filters successively following the same procedure.

\begin{table}[t]
    \centering
    \begin{center}
    \begin{tabular}{@{}lccc@{}}
        \toprule
        & OA & AA & $\lrec$\\
        \midrule
        \textbf{SOL}~\cite{ballet1999studio, cella2020orchideasol} \\
        ~~Direct Classification &  $97.8$ & $94.8$ & --- \\
        ~~APNet~\cite{zinemanas2021deep} & $95.3$ & $91.3$ & $\mathbf{0.1}$\\
        ~~Ours w supervision& $\mathbf{99.3}$ & $\mathbf{95.8}$   & $2.6$\\
        \midrule
        \textbf{LibriSpeech}~\cite{panayotov2015librispeech} \\
        ~~Direct Classification & $99.4$ & $99.5$ & ---\\
        ~~APNet~\cite{zinemanas2021deep} & $97.8$ & $97.8$  & $\mathbf{0.2}$ \\
        ~~Ours w supervision& $\mathbf{99.9}$ & $\mathbf{99.9}$ & $2.6$  \\
        \bottomrule
    \end{tabular}
    \end{center}
    \vspace{-1em}
    \caption{\textbf{Classification Results.}~{Accuracy and reconstruction error computed on the test sets.}}\label{tab:classification}
    \vspace{-1em}
\end{table}

\paragraph{Implementation details.}
We use the ADAM~\cite{kingma2014adam} optimizer with a learning rate of $10^{-4}$, and a weight decay of $10^{-6}$ for the transformation networks and $0$ for $\beta$ and the prototypes. Our model has $545$k parameters with $K=128$ prototypes. For comparison, the reconstruction model proposed by Zinemanas~\etal~\cite{zinemanas2021deep} has $1.8$M parameters.
\section{Experiments}

We evaluate our approach in both supervised and unsupervised settings and on two datasets.

\paragraph{Datasets.}{We consider the following datasets:}
\begin{itemize}[itemsep=-.25em, wide, labelwidth=!, labelindent=0pt, topsep=0cm]
    \item \textbf{SOL~\cite{ballet1999studio, cella2020orchideasol}.}~This dataset contains $24\,450$ samples of individual notes played with various playing techniques by $33$ different instruments and sampled at $44.1$kHz. We evaluate instrument classification.
    \item \textbf{Librispeech~\cite{panayotov2015librispeech}.}~This {1000-hour corpus contains} English speech sampled at $16$kHz. We selected the 128 predominant speakers from the {\tt train-clean-360} set. We evaluate the classification of speaker.
\end{itemize}

{For both datasets, we randomly selected $70\%$ of the clips for training, $10\%$ for validation, and $20\%$ for testing. {We always use as many prototypes as the number of classes in the dataset.}}

\paragraph{Metrics.} To assess the quality of classification models, we report the following metrics:
\begin{itemize}[itemsep=-.25em, wide, labelwidth=!, labelindent=0pt, topsep=0cm]
    \item \textit{Overall Accuracy (OA).}~Percentage of input samples correctly classified by our model, \ie, best reconstructed by the prototype assigned to their true class.
    \item \textit{Average Accuracy (AA).} Average of the classwise accuracy, computed across classes without weights.
    \item \textit{Reconstruction error ($\lrec$).}~{To assess the quality of the reconstruction, we also report the reconstruction error $\lrec$.}
\end{itemize}

Methods trained in the unsupervised setting cluster the audio samples instead of classifying them. We associate each cluster with the majority class of its assigned training samples. We can then predict for each test sample the class of its best-fitting cluster, allowing us to compute all the classification metrics.

\paragraph{Baselines.} To put the performance of our method in context, we trained different baselines:
\begin{itemize}[itemsep=-.25em, wide, labelwidth=!, labelindent=0pt, topsep=0cm]
\item \textit{Classification.}~We trained a temporal convolutional network to classify log-Mel spectrograms. We supervise this network with the cross-entropy on the labels. This network, which does not provide a reconstruction error, is called \emph{Direct Classification}. We also evaluated the APNet~\cite{zinemanas2021deep} approach on our datasets. However, this method is limited in its interpretability, as it relies on latent prototypes and uses a fully connected layer to classify samples based on their similarity to the prototypes.
\item \textit{Clustering.}~We trained a two-dimensional convolutional autoencoder without skip-connections to reconstruct log-Mel spectrograms. We supervised this network with the $\ell_2$ norm between the input and the reconstruction. We then use K-means~\cite{bottou1995convergence} on the innermost feature maps to cluster the samples. We also use K-means directly on the predicted similarities between feature maps and APNet prototypes~\cite{zinemanas2021deep}. Note this last approach requires to first train APNet in a supervised setting.
\end{itemize}

\begin{figure}[t]
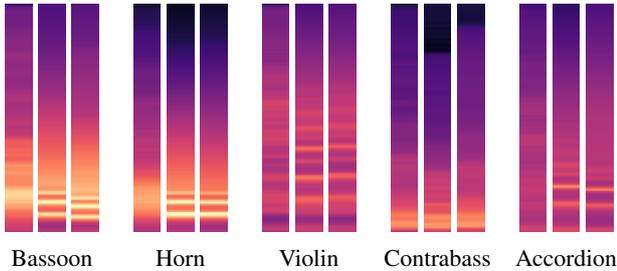

    \centering
    \begin{center}
        \pairspect{2_Bassoons_Bassoon}{Bassoon}
        \pairspect{12_Horns_Horn}{Horn}
        \pairspect{18_Strings_Violin}{Violin}
        \pairspect{20_Strings_Contrabass}{Contrabass}
        \pairspect{29_Keyboards_Accordion}{Accordion}
    \end{center}
    \vspace{-1em}
    \caption{
    \textbf{Learned Meaningful Prototypes.}~{Comparison between the average spectrogram from one class (left), and the prototypes learned in the supervised setting without (middle) and with (right) cross-entropy loss. Our model can learn spectral representations that are crisper and more discriminative than a simple average.}
    }
    \label{fig:protovsmean}
\end{figure}

\subsection{Quantitative Results}\label{sec:quantitativeresults}

\paragraph{Reconstruction.}As can be seen in \tabref{tab:ablation}, each successive addition of a transformation decreases the reconstruction error $\lrec$ and improves the accuracy of classification.
This shows that the curriculum training scheme allows our model to learn insightful and complementary spectral transformations. Supervised with the cross-entropy loss, our model reaches high accuracy, at the cost of a slightly higher reconstruction error. In fact, the additional objective not only encourages the reconstruction models $\mathcal{R}_k$ to provide a good reconstruction for the samples of their assigned class, but also a worse reconstruction for other classes.

Finally, we see that while adding frequency-filters yields better reconstruction, they hurt classification performance on SOL~\cite{ballet1999studio, cella2020orchideasol} in the unsupervised setting. We hypothesize that the filters makes the transformation too expressive, reducing the specificities of the prototypes.

\paragraph{Clustering.} As shown in \tabref{tab:clustering}, our method performs well compared to the baselines. For SOL, our method is affected by the class imbalance between samples, but still produces competitive results. In particular, note that our approach does not need any labels during training,  in contrast to the APNet-based approach~\cite{zinemanas2021deep}. For LibriSpeech, our method produces convincing clustering results for speaker identification.

\paragraph{Classification.}As shown in \tabref{tab:classification}, our method reaches almost perfect precision for both datasets.
While APNet~\cite{zinemanas2021deep} reaches better reconstruction scores, this method learns powerful transformations in an abstract latent space. This leads to a model that is more expressive but less interpretable. In contrast, we restrict the scope of the spectral transformations so that a prototype relates in a meaningful way to the samples that it reconstructs well.
Yet, our model is still capable of producing faithful reconstructions, which would be suitable for audio generation tasks.
\subsection{Interpretability}\label{sec:qualitativeresults}

\begin{figure*}[t]
    \centering
    \newcommand{\reconstructionline}[7]{
    \includegraphics[width=.23\linewidth,height=5em]{images/recs/#1/noerror/input_#2.png} &
    \includegraphics[height=.02\linewidth,width=5em,angle=90]{images/protos/ISMIR_/#1/protos_clu/proto_clu_sample#7.png} &
    \includegraphics[width=.23\linewidth,height=5em]{images/recs/#1/noerror/rec_clu_#3.png} &
    \includegraphics[height=.02\linewidth,width=5em,angle=90]{images/protos/ISMIR_/#1/protos_woce/proto#6.png} &
    \includegraphics[width=.23\linewidth,height=5em]{images/recs/#1/noerror/rec_woce_#4.png} &
    \includegraphics[height=.02\linewidth,width=5em,angle=90]{images/protos/ISMIR_/#1/protos_wce/proto_wce_#6.png} &
    \includegraphics[width=.23\linewidth,height=5em]{images/recs/#1/noerror/rec_wce_#5.png}\\
}
\begin{subfigure}[t]{\textwidth}
    \setlength{\tabcolsep}{1pt}
    \begin{tabular}{@{}ccccccc@{}}
        \multirow{2}{*}{Input sample} & \multicolumn{2}{c}{Unsupervised} & \multicolumn{2}{c}{Supervised w/o $\lce$} & \multicolumn{2}{c}{Supervised w $\lce$} \\\cmidrule(lr){2-3}\cmidrule(lr){4-5}\cmidrule(lr){6-7}
         & $P$ & Reconstruction  & $P$ & Reconstruction & $P$ & Reconstruction\\
        %\reconstructionline{sol}{188_label1}{188_label1}{188_label1}{188_label1}
        %\reconstructionline{sol}{389_label4}{389_label4}{389_label4}{389_label4}
        %\reconstructionline{sol}{698_label10}{698_label10}{698_label10}{698_label10}
        \reconstructionline{sol}{1463_label20}{1463_label20}{1463_label20}{1463_label20}{20_Strings_Contrabass}{0}
        %\reconstructionline{sol}{2399_label30}{2399_label30}{2399_label30}{2399_label30}
        \reconstructionline{sol}{2611_label32}{2611_label32}{2611_label32}{2611_label32}{32_Clarinets_Clarinet-Bb}{1}
    \end{tabular}
    \vspace{-.5em}
    \subcaption{\textbf{SOL~\cite{ballet1999studio, cella2020orchideasol}.}~Reconstructions of an audio clip of contrabass (top) and clarinet (bottom).}
\end{subfigure}
\begin{subfigure}[t]{\textwidth}
    \setlength{\tabcolsep}{1pt}
    \begin{tabular}{@{}ccccccc@{}}
        %Input sample & & Unsupervised & & Supervised w/o $\lce$& & Supervised w $\lce$\\
        \reconstructionline{libri}{1179_label68}{1179_label68}{1179_label68}{1179_label68}{68_6637_F_ChristineNendza}{0}
        \reconstructionline{libri}{1041_label96}{1041_label96}{1041_label96}{1041_label96}{96_581_M_C.Berrius}{1}
    \end{tabular}
    \vspace{-.5em}
    \subcaption{\textbf{LibriSpeech~\cite{panayotov2015librispeech}.}~Reconstructions of an audio clip of C. Nendza (F) (top) and C. Berriux (M) (bottom) speaking.}
\end{subfigure}
\vspace{-.5em}
    \caption{
    \textbf{Reconstruction of Input Sounds.}~{For each input sample, we show the reconstruction provided by different settings of our model as well as the corresponding prototype $P$ used for reconstruction. Note how the timbre of the speaker or instrument is well represented while the inputs fluctuates in pitch and intensity. This leads to an insightful characterization of each instrument or speaker.}}
    \label{fig:reconstructions}
%\end{figure*}
\vspace{1em}
%\begin{figure*}[h]
    \centering
    \newlength{\spectwidth}
    \setlength{\spectwidth}{.135\linewidth}
    \newlength{\spectheight}
    \setlength{\spectheight}{3.3em}
    \newcommand{\presep}{
    %\vspace{-.5em}
    %\rule{0pt}{1em}
}
\newcommand{\postsep}{
    %\vspace{.2em}
    %\rule[-.6em]{0pt}{0pt}
}
    
\setlength{\fboxsep}{0pt}
\setlength{\fboxrule}{2pt}
\newcommand{\boxspect}[2]{
    \fcolorbox{#2}{white}{
        \!\!\includegraphics[width=\spectwidth,height=\spectheight]{#1}\!\!
    }\!\! %\caption*{test}
}
\newcommand{\showinput}[2]{
    \includegraphics[width=\spectwidth,height=\spectheight]{images/recs/sol/input_#1_label#2.png}\!\!
}
\newcommand{\rotquarter}[1]{\rotatebox{90}{#1}}
\newcommand{\showproto}[2]{
    % \raggedleft \bf #1 &
    \includegraphics[width=\spectheight, height=.5em,angle=90]{images/protos/ISMIR_/sol/protos_woce/#2.png}
}
\newcommand{\nameproto}[1]{{{\bf #1}}}

\setlength{\tabcolsep}{2pt}
\renewcommand{\arraystretch}{1}
\begin{tabular}{@{}cc|cccccc@{}}
    & & \multicolumn{6}{c}{\bf Inputs} \\
    & & \bf Harp & \bf Flute & \bf Trumpet-C & \bf Contrabass & \bf Oboe & \bf Clarinet-Bb
    \\
    & & \showinput{188}{1} & \showinput{380}{4} & \showinput{698}{10} & \showinput{1463}{20} & \showinput{2409}{30} & \showinput{2611}{32} %\vspace{-.25em}
    \\\midrule
    %\multirow{12}{*}{\rotquarter{\bf Prototypes}}
    \multirow{6}{*}{\rotatebox{90}{\bf Prototypes~~~~~~~~~~~~~~~~~~~~~~~~~~~~~~~~~~~~~~~}} & \showproto{Harp}{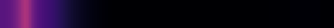} &
    \boxspect{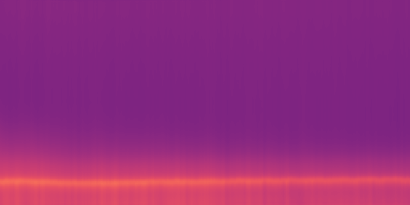}{Kselect} &
    \boxspect{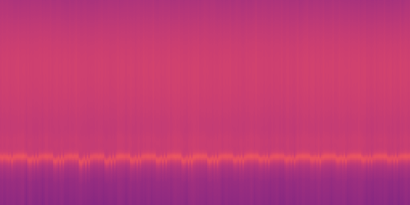}{Knotselect} &
    \boxspect{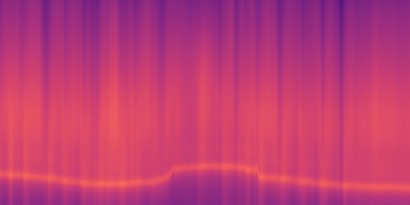}{Knotselect} &
    \boxspect{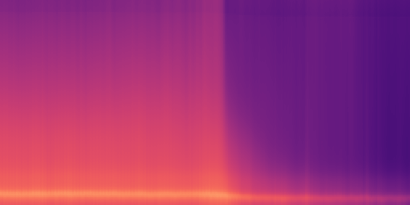}{Knotselect} &
    \boxspect{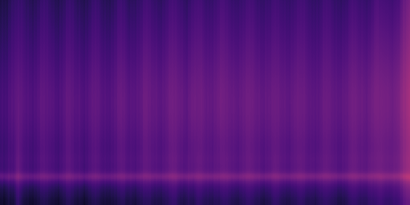}{Knotselect} &
    \boxspect{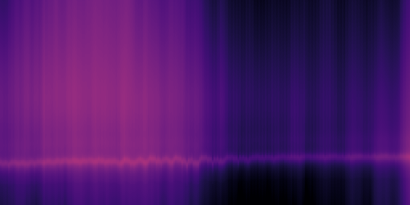}{Knotselect}
    \presep\\
    & \nameproto{Harp} & 
    $\lrec=0.57$ &
    $\lrec=3.39$ &
    $\lrec=3.85$ &
    $\lrec=1.23$ &
    $\lrec=84.32$ &
    $\lrec=69.68$
    \postsep\\
    & \showproto{Flute}{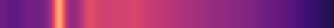} &
    \boxspect{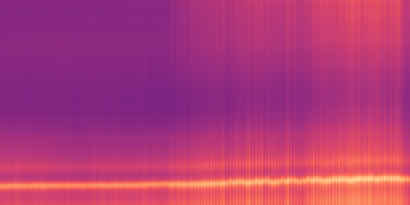}{Knotselect} &
    \boxspect{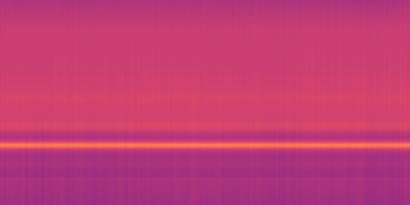}{Kselect} &
    \boxspect{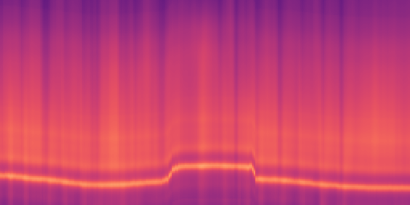}{Knotselect} &
    \boxspect{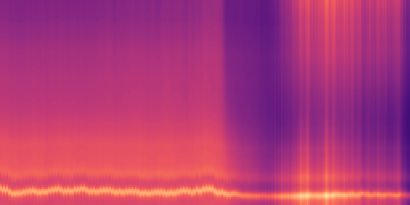}{Knotselect} &
    \boxspect{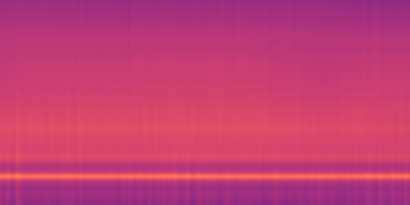}{Knotselect} &
    \boxspect{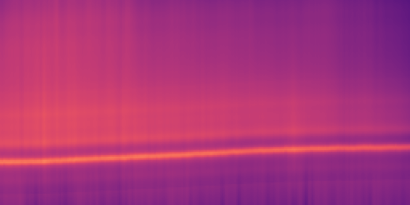}{Knotselect}
    \presep\\
    & \nameproto{Flute} & 
    $\lrec=11.33$ &
    $\lrec=2.74$ &
    $\lrec=2.58$ &
    $\lrec=12.06$ &
    $\lrec=4.80$ &
    $\lrec=5.56$
    \postsep\\
    & \showproto{Trumpet-C}{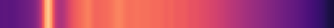} &
    \boxspect{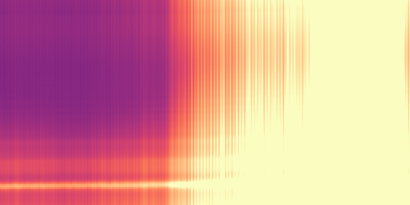}{Knotselect} &
    \boxspect{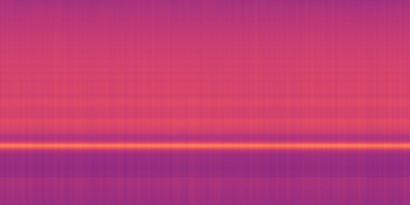}{Knotselect} &
    \boxspect{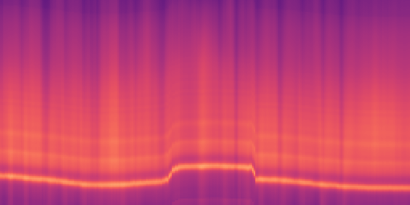}{Kselect} &
    \boxspect{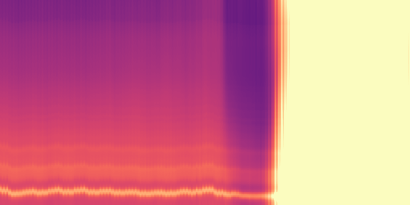}{Knotselect} &
    \boxspect{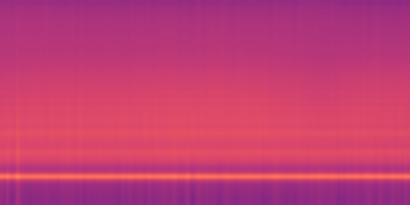}{Knotselect} &
    \boxspect{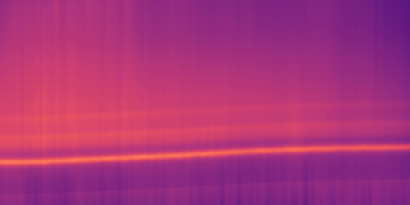}{Knotselect}
    \presep\\
    & \nameproto{Trumpet-C} & 
    $\lrec=264.04$ &
    $\lrec=3.43$ &
    $\lrec=2.49$ &
    $\lrec=392.41$ &
    $\lrec=4.89$ &
    $\lrec=5.32$
    \postsep\\
    & \showproto{Contrabass}{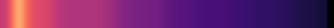} &
    \boxspect{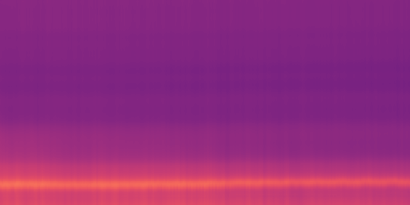}{Knotselect} &
    \boxspect{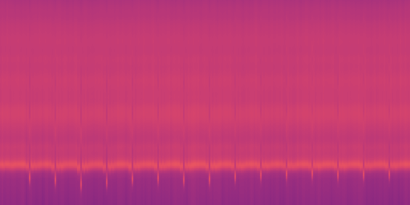}{Knotselect} &
    \boxspect{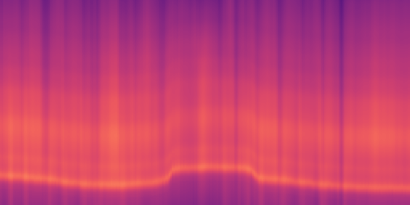}{Knotselect} &
    \boxspect{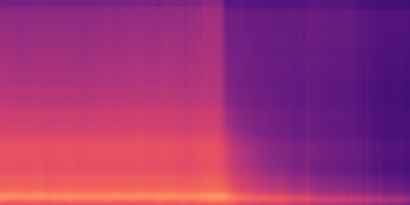}{Kselect} &
    \boxspect{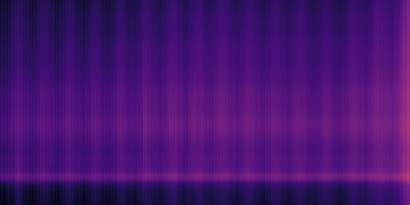}{Knotselect} &
    \boxspect{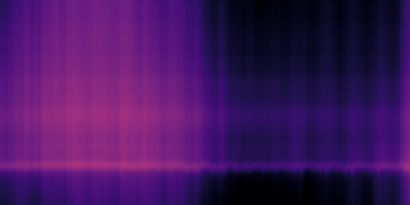}{Knotselect}
    \presep\\
    & \nameproto{Contrabass} & 
    $\lrec=0.70$ &
    $\lrec=4.33$ &
    $\lrec=2.88$ &
    $\lrec=1.01$ &
    $\lrec=101.53$ &
    $\lrec=80.49$
    \postsep\\
    & \showproto{Oboe}{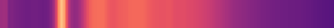} &
    \boxspect{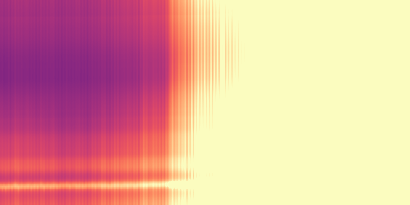}{Knotselect} &
    \boxspect{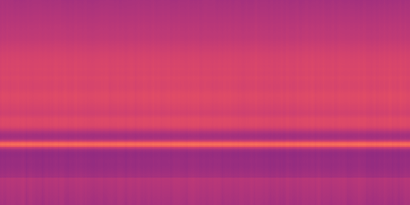}{Knotselect} &
    \boxspect{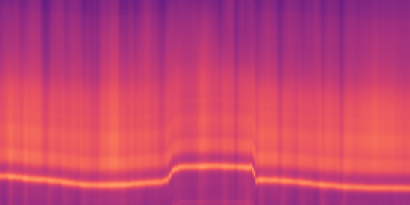}{Knotselect} &
    \boxspect{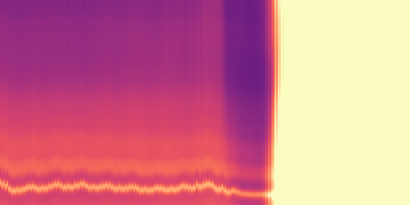}{Knotselect} &
    \boxspect{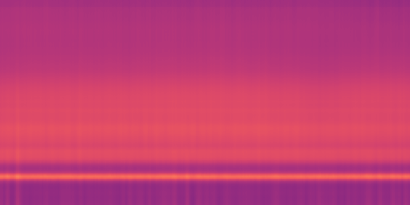}{Kselect} &
    \boxspect{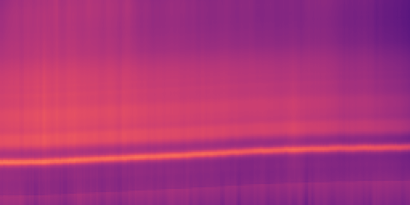}{Knotselect}
    \presep\\
    & \nameproto{Oboe} & 
    $\lrec=410.58$ &
    $\lrec=3.92$ &
    $\lrec=2.71$ &
    $\lrec=474.13$ &
    $\lrec=4.56$ &
    $\lrec=5.51$
    \postsep\\
    & \showproto{Clarinet-Bb}{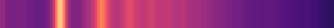} &
    \boxspect{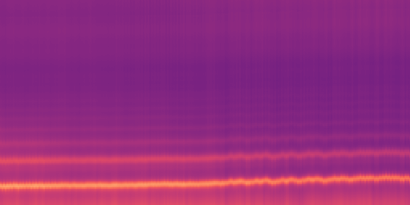}{Knotselect} &
    \boxspect{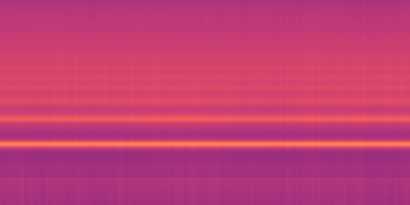}{Knotselect} &
    \boxspect{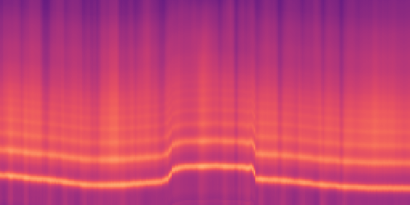}{Knotselect} &
    \boxspect{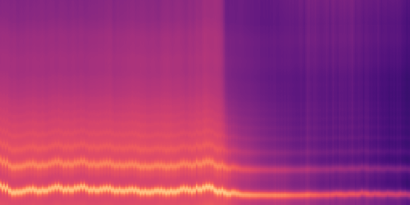}{Knotselect} &
    \boxspect{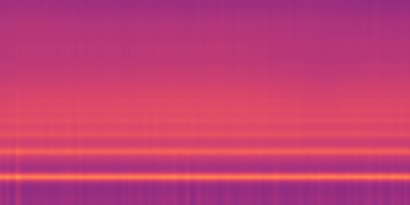}{Knotselect} &
    \boxspect{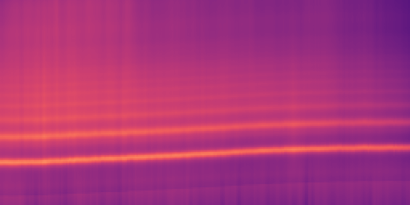}{Kselect}
    \presep\\
    & \nameproto{Clarinet-Bb} & 
    $\lrec=2.04$ &
    $\lrec=3.40$ &
    $\lrec=2.85$ &
    $\lrec=2.80$ &
    $\lrec=5.50$ &
    $\lrec=4.41$\\
\end{tabular}
\vspace{-.5em}
    \caption{
    \textbf{Reconstructions from Different Prototypes.}~{We show the reconstructions of input samples (columns) from SOL~\cite{ballet1999studio, cella2020orchideasol} by learned prototypes (lines) in the supervised setting without cross-entropy loss. The prototype of the corresponding instruments provides the lowest reconstruction error (green frame).}}
    \label{fig:transfer}
\end{figure*}

As shown in \figref{fig:reconstructions}, our model learns to reconstruct complex input samples (\eg different notes, techniques, or voices) by automatically adjusting the amplitude, pitch, and spectral support of spectral prototypes. 
As seen on \figref{fig:protovsmean} and \figref{fig:teaser}, these prototypes capture rich spectral characteristics such as timbre, and can be interpreted as an ``acoustical fingerprint''. Furthermore, as the prototypes are given in the spectral domain, they can be easily played and analyzed. Audio examples featuring the learned prototypes are provided in the supplementary materials.

In \figref{fig:transfer}, we represent the reconstruction of sounds played by various instrument with different prototypes. While the reconstructions are clearly more faithful when using the instruments' dedicated prototype, this opens the perspective for further MIR application such as interpretable timbre transfer.

\section{Conclusion}

We presented a novel approach to audio understanding by representing large collections of audio clips with few spectral prototypes equipped with learned transformation networks. Our model produces concise, expressive, and interpretable representations of raw audio excerpts that can be used to obtain state-of-the-art results for audio classification and clustering tasks.

\paragraph{Acknowledgements.}{This work was supported in part by ANR project READY3D ANR-19-CE23-0007 and was granted access to the HPC resources of IDRIS under the allocation 2022-AD011012096R1 made by GENCI. We thank Theo Deprelle, Nicolas Gonthier, Tom Monnier and  Yannis Siglidis for inspiring discussions and feedbacks.
}

\clearpage
% For bibtex users:
\bibliography{main.bib}

\begin{thebibliography}{10}
\providecommand{\url}[1]{#1}
\csname url@samestyle\endcsname
\providecommand{\newblock}{\relax}
\providecommand{\bibinfo}[2]{#2}
\providecommand{\BIBentrySTDinterwordspacing}{\spaceskip=0pt\relax}
\providecommand{\BIBentryALTinterwordstretchfactor}{4}
\providecommand{\BIBentryALTinterwordspacing}{\spaceskip=\fontdimen2\font plus
\BIBentryALTinterwordstretchfactor\fontdimen3\font minus
  \fontdimen4\font\relax}
\providecommand{\BIBforeignlanguage}[2]{{%
\expandafter\ifx\csname l@#1\endcsname\relax
\typeout{** WARNING: IEEEtran.bst: No hyphenation pattern has been}%
\typeout{** loaded for the language `#1'. Using the pattern for}%
\typeout{** the default language instead.}%
\else
\language=\csname l@#1\endcsname
\fi
#2}}
\providecommand{\BIBdecl}{\relax}
\BIBdecl

\bibitem{abesser2020review}
J.~Abe{\ss}er, ``A review of deep learning based methods for acoustic scene
  classification,'' \emph{Applied Sciences}, 2020.

\bibitem{ndou2021music}
N.~Ndou, R.~Ajoodha, and A.~Jadhav, ``Music genre classification: A review of
  deep-learning and traditional machine-learning approaches,'' in
  \emph{IEMTRONICS}, 2021.

\bibitem{monnier2020deep}
T.~Monnier, T.~Groueix, and M.~Aubry, ``{Deep Transformation-Invariant
  Clustering},'' in \emph{NeurIPS}, 2020.

\bibitem{loiseau2021representing}
R.~Loiseau, T.~Monnier, M.~Aubry, and L.~Landrieu, ``{Representing Shape
  Collections with Alignment-Aware Linear Models},'' in \emph{3DV}, 2021.

\bibitem{jaderberg2015spatial}
M.~Jaderberg, K.~Simonyan, and A.~Zisserman, ``Spatial transformer networks,''
  \emph{NeurIPS}, 2015.

\bibitem{ballet1999studio}
G.~Ballet, R.~Borghesi, P.~Hoffmann, and F.~L{\'e}vy, ``Studio online 3.0: An
  internet "killer application" for remote access to {IRCAM} sounds and
  processing tools,'' in \emph{Journ{\'e}es d'Informatique Musicale}, 1999.

\bibitem{cella2020orchideasol}
C.~E. Cella, D.~Ghisi, V.~Lostanlen, F.~L{\'e}vy, J.~Fineberg, and Y.~Maresz,
  ``{OrchideaSOL:} a dataset of extended instrumental techniques for
  computer-aided orchestration,'' \emph{ICMC}, 2020.

\bibitem{panayotov2015librispeech}
V.~Panayotov, G.~Chen, D.~Povey, and S.~Khudanpur, ``Librispeech: an asr corpus
  based on public domain audio books,'' in \emph{ICASSP}, 2015.

\bibitem{bhalke2016automatic}
D.~Bhalke, C.~Rao, and D.~S. Bormane, ``Automatic musical instrument
  classification using fractional fourier transform based-{MFCC} features and
  counter propagation neural network,'' \emph{Journal of Intelligent
  Information Systems}, 2016.

\bibitem{lostanlen2018extended}
V.~Lostanlen, J.~And{\'e}n, and M.~Lagrange, ``Extended playing techniques: the
  next milestone in musical instrument recognition,'' in \emph{International
  Conference on Digital Libraries for Musicology}, 2018.

\bibitem{bai2021speaker}
Z.~Bai and X.-L. Zhang, ``Speaker recognition based on deep learning: An
  overview,'' \emph{Neural Networks}, 2021.

\bibitem{ye2021deep}
F.~Ye and J.~Yang, ``A deep neural network model for speaker identification,''
  \emph{Applied Sciences}, 2021.

\bibitem{tits2019visualization}
N.~Tits, F.~Wang, K.~E. Haddad, V.~Pagel, and T.~Dutoit, ``Visualization and
  interpretation of latent spaces for controlling expressive speech synthesis
  through audio analysis,'' \emph{Conference of the International Speech
  Communication Association}, 2019.

\bibitem{roche2018autoencoders}
F.~Roche, T.~Hueber, S.~Limier, and L.~Girin, ``Autoencoders for music sound
  modeling: a comparison of linear, shallow, deep, recurrent and variational
  models,'' \emph{Sound and Music Computing}, 2018.

\bibitem{naranjo2020open}
J.~Naranjo-Alcazar, S.~Perez-Castanos, P.~Zuccarello, F.~Antonacci, and
  M.~Cobos, ``Open set audio classification using autoencoders trained on few
  data,'' \emph{Sensors}, 2020.

\bibitem{li2018deep}
O.~Li, H.~Liu, C.~Chen, and C.~Rudin, ``Deep learning for case-based reasoning
  through prototypes: A neural network that explains its predictions,'' in
  \emph{AAAI}, 2018.

\bibitem{zinemanas2021deep}
P.~Zinemanas, M.~Rocamora, M.~Miron, F.~Font, and X.~Serra, ``An interpretable
  deep learning model for automatic sound classification,'' \emph{Electronics},
  2021.

\bibitem{monnier2021dtisprites}
T.~Monnier, E.~Vincent, J.~Ponce, and M.~Aubry, ``{Unsupervised Layered Image
  Decomposition into Object Prototypes},'' in \emph{ICCV}, 2021.

\bibitem{stevens1937scale}
S.~S. Stevens, J.~Volkmann, and E.~B. Newman, ``A scale for the measurement of
  the psychological magnitude pitch,'' \emph{The Journal of the Acoustical
  Society of America}, 1937.

\bibitem{tamaru2019generative}
H.~Tamaru, Y.~Saito, S.~Takamichi, T.~Koriyama, and H.~Saruwatari, ``Generative
  moment matching network-based random modulation post-filter for dnn-based
  singing voice synthesis and neural double-tracking,'' in \emph{ICASSP}, 2019.

\bibitem{andreux2018music}
M.~Andreux and S.~Mallat, ``Music generation and transformation with moment
  matching-scattering inverse networks,'' in \emph{ISMIR}, 2018.

\bibitem{bottou1995convergence}
L.~Bottou and Y.~Bengio, ``Convergence properties of the k-means algorithms,''
  in \emph{NeurIPS}, 1995.

\bibitem{ronneberger2015u}
O.~Ronneberger, P.~Fischer, and T.~Brox, ``U-net: Convolutional networks for
  biomedical image segmentation,'' \emph{MICCAI}, 2015.

\bibitem{kingma2014adam}
D.~P. Kingma and J.~Ba, ``Adam: A method for stochastic optimization,''
  \emph{ICLR}, 2015.

\end{thebibliography}

\end{document}